\documentclass{appolb}

\usepackage{amsmath,amssymb,amsopn}
\usepackage{array}
\usepackage{feynmf}
\usepackage{graphicx}

\def\lg{{\mathchoice{~\raise.58ex\hbox{$<$}\mkern-14.8mu\lower.52ex\hbox{$>$}~}
                    {~\raise.58ex\hbox{$<$}\mkern-14.8mu\lower.52ex\hbox{$>$}~}
                    {\raise.59ex\hbox{{$\scriptscriptstyle <$}}\mkern-12.8mu%
                     \lower.01ex\hbox{{$\scriptscriptstyle >$}}}   {}   }}
\def\gl{{\mathchoice{~\raise.58ex\hbox{$>$}\mkern-12.8mu\lower.52ex\hbox{$<$}~}
                    {~\raise.58ex\hbox{$>$}\mkern-12.8mu\lower.52ex\hbox{$<$}~}
                    {\raise.62ex\hbox{{$\scriptscriptstyle >$}}\mkern-12.0mu%
                     \lower.05ex\hbox{{$\scriptscriptstyle <$}}}  {}    }}

\newcommand{\be}{\begin{equation}}
\newcommand{\ee}{\end{equation}}
\newcommand{\ba}{\begin{eqnarray}}
\newcommand{\ea}{\end{eqnarray}}
\newcommand{\ban}{\begin{eqnarray*}}
\newcommand{\ean}{\end{eqnarray*}}
\newcommand \nn {\nonumber}
\newcommand{\sla}{\!\!\!/ \,}

\begin{document}

\title{Supersymmetric QED Plasma\thanks{Presented by A. Czajka at HIC-for-FAIR 
Workshop \& XXVIII Max Born Symposium, Wroc\l aw, Poland, May 19-21, 2011. }
}
\author{Alina Czajka
\address{Institute of Physics, Jan Kochanowski University, Kielce, Poland}
\and
Stanis\l aw Mr\' owczy\' nski
\address{Institute of Physics, Jan Kochanowski University, Kielce, Poland \\
and National Centre  for Nuclear Research, Warsaw, Poland}
}
\date{October 26, 2011}
\maketitle

\begin{abstract}

We systematically compare the ${\cal N} =1$ SUSY QED plasma to its non-supersymmetric 
counterpart which is QED plasma of electrons, positrons and photons. Collective excitations 
and collisional processes in the two systems are confronted to each other in a regime of small 
coupling.  The collective and collisional characteristics of supersymmetric plasma 
are both very similar to those of QED plasma.

\end{abstract}

%%%%%%%%%%%%%%%%%%%%%%%%%%%%%%%%%%%%%%%%%%%%%
\section{Introduction}
%%%%%%%%%%%%%%%%%%%%%%%%%%%%%%%%%%%%%%%%%%%%%

Supersymmetry is a good candidate to be a symmetry of Nature at sufficiently high energies 
and if true, SUSY plasmas existed in the early Universe. Experiments at the Large 
Hadron Collider can soon provide an evidence of supersymmetry. However, independently 
of its ontological status, supersymmetric field theories are worth studying because of their 
unique features. A discovery of the AdS/CFT duality of  the five-dimensional gravity in  
anti de Sitter geometry and the conformal field theories, see the review \cite{Aharony:1999ti}, 
stimulated a great interest in the ${\cal N} = 4$ supersymmetric Yang-Mills theory which is both 
classically and quantum mechanically conformally invariant. The duality offers a unique tool to 
study strongly coupled field theories,  as the gravitational constant is inversely proportional to 
the coupling constant  of dual conformal field theory and thus some problems of strongly 
coupled field theories can be solved via weakly coupled gravity. Some intriguing results have 
been obtained, see the reviews \cite{Son:2007vk,Janik:2010we},  but relevance of the results 
for non-supersymmetric theories, which are of our actual interest, remains an open issue.
One asks how properties of the plasma governed by ${\cal N} =4$ Super Yang-Mills theory 
are related to those of the usual quark-gluon plasma experimentally studied in relativistic 
heavy-ion collisions. While such a comparison of the two systems is a difficult task, some 
comparative analyses have been done in the domain of weak coupling where perturbative 
methods are applicable \cite{CaronHuot:2006te,CaronHuot:2008uh,Blaizot:2006tk,Chesler:2006gr,Chesler:2009yg}. 

Our aim is to systematically compare the ${\cal N} =4$ Super Yang-Mills plasma to the
QCD one with a particular emphasis on non-equilibrium characteristics which has not
attracted much attention yet.  We started, however, with the supersymmetric  ${\cal N} =1$ 
QED plasma which is noticeably simpler than that of ${\cal N} =4$ Super Yang-Mills.
We first studied collective excitations of ultrarelativistic ${\cal N} =1$ SUSY QED  plasma 
which were confronted with those of an electromagnetic plasma of electrons, positrons and photons
\cite{Czajka:2010zh}. In the subsequent paper \cite{Czajka:2011zn}, we focused on collisional 
processes which control plasma transport properties. Here we summarize our all findings. 
Throughout the paper we use a  natural system of units with 
$c= \hbar = k_B =1$ and the metric tensor of the signature $(+ - - -)$.

%%%%%%%%%%%%%%%%%%%%%%%%%%%%%%%%%%%%%%%%%%%%%
\section{${\cal N}=1$ SUSY QED}
\label{sec-lagrangian}
%%%%%%%%%%%%%%%%%%%%%%%%%%%%%%%%%%%%%%%%%%%%%

We start our considerations by writing down the Lagrangian of ${\cal N}=1$ SUSY QED
which is known, see {\it e.g.} \cite{Binoth:2002xg}, to be
\ba
{\cal L} &=& -\frac{1}{4}F^{\mu \nu} F_{\mu \nu} +\frac{i}{2} \bar \Lambda \partial \sla \Lambda
+  i\bar \Psi D\!\sla \Psi
+(D_\mu \phi_L)^*(D^\mu \phi_L) + (D_\mu^* \phi_R)(D^\mu \phi_R^*) \nn
\\ \nn
&& +\sqrt{2} e \big( \bar \Psi P_R \lambda \phi_L - \bar \Psi P_L \lambda \phi_R^*
+ \phi_L^* \bar \Lambda P_L \Psi - \phi_R \bar \Lambda P_R \Psi \big) \\
&& -\frac{e^2}{2} \big( \phi_L^* \phi_L - \phi_R^* \phi_R \big)^2 ,
\ea
where the strength tensor $F^{\mu \nu}$ is expressed through the electromagnetic
four-potential $A^\mu$ as $F^{\mu \nu} \equiv \partial^\mu A^\nu - \partial^\nu A^\mu$
and the covariant derivative is $D^\mu \equiv \partial^\mu +ie A^\mu$;
$\Lambda$ is the Majorana bispinor photino field, $\Psi$ is the
Dirac bispinor electron field, $\phi_L$ and $\phi_R$ are the scalar left selectron
and right selectron fields; the projectors $P_L$ and $P_R$ are defined in a standard
way $P_L \equiv \frac{1}{2}(1 - \gamma_5)$ and $P_R \equiv \frac{1}{2}(1 + \gamma_5)$.
Since we are interested in ultrarelativistic plasmas, the mass terms are neglected
in the Lagrangian.

%%%%%%%%%%%%%%%%%%%%%%%%%%%%%%%%%%%%%%%%%%%%%
\section{Collective excitations}
\label{sec-collective}
%%%%%%%%%%%%%%%%%%%%%%%%%%%%%%%%%%%%%%%%%%%%%

We consider collective excitations of ${\cal N}=1$ SUSY QED plasma which is 
homogeneous but the momentum distribution is, in general, different from the equilibrium one. 
The excitations are determined by the dispersion equations which for (quasi-)photons, 
electrons, photinos and selectrons read
\ba
\label{dis-photon}
{\rm det}\Big[ k^2 g^{\mu \nu} -k^{\mu} k^{\nu} - \Pi^{\mu \nu}(k) \Big]
 = 0 ,
\\
\label{dis-electron}
 {\rm det}\Big[ k\sla  - \Sigma (k) \Big]  = 0,
\\
\label{dis-photino}
{\rm det}\Big[ k\sla  - \tilde \Pi (k) \Big] = 0 ,
\\
\label{dis-selectron}
k^2 + \tilde\Sigma_{L,R} (k) = 0,
\ea
where $\Pi^{\mu \nu}(k), \; \Sigma (k), \; \tilde \Pi(k), \; \tilde\Sigma_{L,R} (k)$ are the retarded 
self-energies of photons, electrons, photinos and  left or right selectrons. Since, the plasma 
under consideration is generally out of equilibrium, the self energies were computed using 
the Keldysh-Schwinger formalism. We were interested in collective modes which occur, 
when wavelength of a quasi-particle is much bigger than a characteristic interparticle 
distance in the plasma, therefore we worked in the Hard Loop Approach, see the review
\cite{Kraemmer:2003gd}, which had been generalized to anisotropic  systems in
\cite{Mrowczynski:2000ed,Mrowczynski:2004kv}. We also assumed that the plasma is 
electrically neutral and that the distribution functions of particles and antiparticles are equal 
to each other. The distribution functions of left and right selectrons are assumed to be the 
same as well.

Self-energies are usually defined by means of a Dyson-Schwinger equation which for the case
of polarization tensor $\Pi$ has the following symbolic form ${\cal D} = D - D \,\Pi  \,{\cal D} $,
where ${\cal D}$ and $D$ is the interacting and free photon propagator, respectively. We 
computed the self energies perturbatively at one-loop level. Within the Keldysh-Schwinger formalism 
one first finds the contour self-energies and further on the retarded self-energies is extracted. 
Using the Hard Loop Approximation, the polarization tensor was found as
\ba
\label{Pi-k-final}
\Pi^{\mu \nu}(k) &=&  4e^2  \int \frac{d^3p}{(2\pi )^3} \;
\frac{f_e({\bf p})+f_s({\bf p})}{E_p} 
\\ \nn
&\times&
 \frac{k^2 p^\mu p^\nu - \big(p^\mu k^\nu + k^\mu p^\nu - g^{\mu \nu} (k \cdot p)\big)(k \cdot p)}
{(k \cdot p + i0^+)^2},
\ea
where $f_e$ and $f_s$ are the distribution functions of electrons and selectrons. 
As seen, $\Pi$ vanishes in the vacuum limit $(f_e , f_s \rightarrow 0)$ which is
a genuine feature of supersymmetric theories. In the non-supersymmetric counterpart,
the vacuum contribution to $\Pi$ diverges and it requires a special treatment. 

Up to the vacuum contribution, the polarization tensor of supersymmetric plasma and of its 
non-supersymmetric counterpart have the same structure.  Therefore, the spectra of collective 
excitations of gauge bosons in the two systems are identical. In equilibrium plasma we have
the longitudinal (plasmon) mode and the transverse one which are discussed in {\it e.g.} the 
textbook \cite{lebellac}. When the plasma is out of equilibrium there is a whole variety of possible 
collective excitations. In particular, there are unstable modes, see  {\it e.g.} the review  \cite{Mrowczynski:2007hb}, which exponentially grow in time and strongly influence the system's 
dynamics.

The one-loop electron self-energy, which was found as
\ba
\label{Si-k-final}
\Sigma (k) &=& e^2  \int \frac{d^3p}{(2\pi )^3} \;
 \frac{ f_\gamma ({\bf p}) +  f_e ({\bf p}) + f_{\tilde \gamma} ({\bf p})  +  f_s ({\bf p})}{E_p}  \,
\frac{ p\sla }{k\cdot p + i 0^+} ,
\ea
with $f_\gamma$ and $f_{\tilde \gamma}$ being the distribution functions of photons and
photinos, has the same structure for the SUSY QED and QED plasma. Therefore, we have 
identical spectrum of excitations of charged fermions in the two systems. In equilibrium 
plasma there are two modes, see in {\it e.g.} the textbook \cite{lebellac}, of opposite helicity
over chirality ratio.  One mode corresponds to the positive energy fermion, another one, 
sometimes called a plasmino, is a specific medium effect.  Out of equilibrium the spectrum
changes but no unstable modes have been found even for an extremely  anisotropic momentum 
distribution  \cite{Mrowczynski:2001az,Schenke:2006fz}.

The photino self-energy was computed  as 
\be
\label{Pi-k-f-final}
\tilde \Pi(k) = e^2 \int \frac{d^3p}{(2\pi )^3} \;
\frac{f_s({\bf p})+f_e({\bf p})}{E_p} \; \frac{ p\sla }{k\cdot p + i 0^+} 
\ee
and its structure coincides with the electron self-energy (\ref{Si-k-final}). The spectra of collective 
excitations are also identical.  When the plasma momentum distribution is anisotropic and unstable 
photon modes occur, the photino modes remain stable. Therefore, the supersymmetry  does not 
induce an instability in the photino sector, as one could naively expect.

Finally, we present the one-loop retarded self-energy of selectron 
\be
\label{Si-k-r-final}
\tilde \Sigma (k) = -2 e^2 \int \frac{d^3p}{(2\pi )^3} \,
\frac{f_e ({\bf p}) + f_\gamma ({\bf p}) +  f_s ({\bf p})+  f_{\tilde \gamma} ({\bf p}) }{E_p},
\ee
which is the same for left and right selectron fields.  Because of supersymmetry  it vanishes in the 
vacuum limit when all the distributions functions are zero. This is also effect of supersymmetry  
that the distribution functions of electrons and selectrons and of photons and photions enter 
Eq.~(\ref{Si-k-r-final}) with the same coefficients. 

The selectron self-energy (\ref{Si-k-r-final})  is independent of $k$, it is negative and real. Therefore,  
$\tilde \Sigma $ can be written as $\tilde \Sigma = - m^2_{\rm eff}$ where $m_{\rm eff}$ 
is the effective selectron mass. Then, the solutions of dispersion equation (\ref{dis-selectron}) are 
$E_k = \pm \sqrt{m^2_{\rm eff} + {\bf k}^2}$.

%%%%%%%%%%%%%%%%%%%%%%%%%%%%%%%%%%%%%%%%%%%%%
\section{Effective Action}
\label{sec-eff-action}
%%%%%%%%%%%%%%%%%%%%%%%%%%%%%%%%%%%%%%%%%%%%%

The Hard Loop Approach can be formulated in an elegant and compact way by using the 
effective action, the form of which is dictated by structure of self energies.  Since the self energy 
of a given field is the second functional derivative of the action with respect to the field, the action
of, say, electromagnetic field is of the form
\be
\label{action-A-1}
{\cal L}^{(A)}_2(x) =
\frac{1}{2} \int d^4y \; A_\mu(x) \Pi^{\mu \nu}(x-y) A_\nu(y) ,
\ee
where the polarization tensor is given by Eq.~(\ref{Pi-k-final}). The subscript `2' indicates that 
the above effective action generates only the two-point function. To generate $n-$point functions 
the action ${\cal L}_2$ needs to be extended to a gauge invariant form by replacing the ordinary 
derivatives by the covariants. Repeating the calculations described in detail in 
\cite{Mrowczynski:2004kv}, one finds the Hard Loop effective actions of ${\cal N}=1$ SUSY QED as
\ba
\label{action-A-2}
{\cal L}^{(A)}_{\rm HL} &=&
4e^2  \int \frac{d^3p}{(2\pi )^3} \,
\frac{f_1({\bf p})}{E_p} \,
F_{\mu \nu} (x) {p^\nu p^\rho \over (p \cdot \partial)^2} F_\rho^{\;\;\mu} (x) ,
\\ [2mm]
\label{action-Psi-2}
{\cal L}^{(\Psi)}_{\rm HL} &=&   i e^2
\int \frac{d^3p}{(2\pi )^3} \; \frac{ f_2 ({\bf p}) }{E_p}  \,
\bar{\Psi}(x) {p \cdot \gamma \over p\cdot D} \Psi (x) ,
\\ [2mm]
\label{action-Lambda-2}
{\cal L}^{(\Lambda)}_{\rm HL} &=&
i e^2 \int \frac{d^3p}{(2\pi )^3} \;
\frac{f_1({\bf p})}{E_p} \;
\bar\Lambda (x) {p \cdot \gamma \over p\cdot \partial} \Lambda (y) ,
\\ [2mm]
\label{action-Phi-2}
{\cal L}^{(\phi_{L,R})}_{\rm HL} &=&  -
2 e^2 \int \frac{d^3p}{(2\pi )^3} \,
\frac{f_2 ({\bf p})}{E_p}
\; \phi_{L,R}^*(x) \phi_{L,R} (x) ,
\ea
where $f_1 ({\bf p})  \equiv f_e ({\bf p}) +  f_s ({\bf p})$ and 
$f_2 ({\bf p}) \equiv f_e ({\bf p}) + f_\gamma ({\bf p}) +  f_s ({\bf p}) +  f_{\tilde \gamma} ({\bf p})$.

The actions  (\ref{action-A-2}, \ref{action-Psi-2}, \ref{action-Lambda-2}, \ref{action-Phi-2})
are obtained from the self-energies but the reasoning can be turned around. As argued
in \cite{Frenkel:1991ts,Braaten:1991gm}, the actions of gauge bosons (\ref{action-A-2}),
charged fermions (\ref{action-Psi-2}) and charged scalars (\ref{action-Phi-2}) are of
unique gauge invariant form.  Therefore, the respective self-energies can be, in principle,
inferred from the known QED self-energies with some help of supersymmetry arguments.
An explicit computation of photino self-energy seems to be unavoidable.

%%%%%%%%%%%%%%%%%%%%%%%%%%%%%%%%%%%%%%%%%%%%%
\section{Collisional processes}
\label{sec-collisions}
%%%%%%%%%%%%%%%%%%%%%%%%%%%%%%%%%%%%%%%%%%%%%

As seen in the lagrangian (\ref{sec-lagrangian}), there is a self-interaction of selectron field
due to the terms $ ( \phi_L^* \phi_L )^2$, $( \phi_R^* \phi_R)^2$ and
$-2 \phi_L^* \phi_L \phi_R^* \phi_R$, there is also a four-boson coupling
$ \phi_{L,R}^* \phi_{L,R} A^\mu A_\mu$. Such a contact interaction is qualitatively different 
than that caused by an exchange of massless boson. The scattering cross section in the absence 
of other interactions is isotropic in the center-of-mass frame of colliding particles and the energy 
and momentum transfers are bigger than that in electromagnetic interactions. Therefore, one 
expects that transport  properties of supersymmetric ${\cal N} =1$ QED plasma differ from 
those of  QED plasma of electrons, positrons and photons. To test this expectation, we computed 
the cross sections of all binary processes which occur in ${\cal N} =1$ SUSY QED  plasma in 
the lowest non-trivial order of $\alpha \equiv e^2/4\pi$. The complete list of processes is 
presented in \cite{Czajka:2011zn}.

There are five processes where only electrons, positrons and photons take part. These processes 
occur in both the supersymmetric QED plasma and usual electromagnetic one. There are twenty 
eight other processes which are characteristic for the  ${\cal N} =1$ SUSY QED. Among those 
processes there are eight of special interest, Compton scattering of selectrons is an example, 
the cross section of which is independent of momentum transfer $t$ or $u$.  The matrix element 
of such a process is simply a number. These processes are qualitatively different from those in 
electromagnetic plasmas which are dominated by an interaction with small momentum transfer. 
We note that for each plasma particle $e$, $\gamma$, $\tilde{e}$, $\tilde{\gamma}$ such a process 
exists. 

It should be remembered that the temperature $T$ is the only dimensional parameter which characterizes 
an equilibrium ultrarelativistic plasma. Consequently, the parametric form of transport coefficients can be 
determined by dimensional arguments. For example, the shear viscosity must be proportional to 
$T^3/\alpha^2$ and it is thus hard to expect that the viscosity of supersymmetric plasma is qualitatively 
different than that of electromagnetic one. Indeed, the shear viscosity of an ${\cal N} =4$ Super Yang-Mills 
plasma is rather similar to that of a quark-gluon plasma \cite{Huot:2006ys}. 

We considered two transport characteristics of the ${\cal N} =1$ QED plasma which are not so constrained 
by dimensional arguments. Specifically, we computed the collisional energy loss and momentum broadening 
of a particle traversing the equilibrium plasma. The dimensional argument does not work here because the
two quantities depend not only on the plasma temperature $T$ but on the test particle energy $E$ as well.

When the matrix element equals $|\mathcal{M}|^2 = 4 e^4$, as in the case of scattering of selectron on photons, 
the energy loss of high energy particle $(E \gg T)$ in equilibrium  ${\cal N} =1$ SUSY QED plasma was found 
to be 
\be
\label{e-loss-BigE}
\frac{d E}{d x}=   -\frac{e^4}{2^5 3 \pi}\, T^2 ,
\ee
which should be confronted with the energy loss of an energetic muon in ultrarelativistic QED 
plasma of electrons, positrons and photons \cite{Braaten:1991jj}
\be
\label{e-loss-QED}
\frac{dE}{dx}= -\frac{e^4}{48\pi^3}\, T^2 \Big( \ln\frac{E}{eT}+2.031\Big).
\ee
As seen, the formulas (\ref{e-loss-BigE}, \ref{e-loss-QED}) are similar to each other up the logarithm term 
discussed in \cite{Czajka:2011zn}. The similarity is rather surprising if one realizes how different are the differential 
cross sections of interest.  

The energy loss can be estimated as $\frac{dE}{d x} \sim \langle \Delta E \rangle / \lambda$, 
where $\langle \Delta E \rangle$ is the typical change of particle's energy in a single collision 
and $\lambda$ is the particle's mean free path given as $\lambda^{-1} = \rho \, \sigma$ with 
$\rho \sim T^3$ being the density of scatterers and $\sigma$ denoting the cross section. For 
the differential cross section $\frac{ d\sigma}{dt} \sim e^4/s^2$, the total cross section is 
$\sigma \sim e^4/s$. When a highly energetic particle with energy $E$ scatters on massless 
plasma particle, $s \sim ET$ and consequently  $\sigma \sim e^4/(ET)$. The inverse mean 
free path is thus estimated as $\lambda^{-1} \sim e^4 T^2/E$.  When the scattering process 
is independent of momentum transfer, $\langle \Delta E \rangle$ is of order $E$ and we finally 
find $-\frac{dE}{d x} \sim e^4 T^2$. When compared to the case of Coulomb scattering, the 
energy transfer in a single collision is much bigger but the cross section is smaller in the 
same proportion. Consequently, the two interactions corresponding to very different differential 
cross sections lead to very similar energy losses.  

We also computed  the momentum broadening, which is usually denoted as $\hat q$,  due to 
the scattering which is momentum-transfer independent. $\hat q$ determines a magnitude of 
radiative energy loss of a highly energetic particle in a plasma medium \cite{Baier:1996sk}. When 
the matrix element equals $|\mathcal{M}|^2 = 4 e^4$, the momentum broadening of a highly 
energetic particle is 
\be
\label{qhat-final}
\hat{q} = \frac{e^4 \zeta(3)}{12 \pi^3}\, T^3
\ee
and it should be compared to the momentum broadening driven by one-photon exchange which is 
of order $e^4 \ln(1/e) \, T^3$ \cite{Arnold:2008vd}. As seen, the momentum broadening and 
consequently the radiative energy loss of a highly energetic particle in SUSY QED and QED 
plasma are similar (up to the logarithm term) to each other.

%%%%%%%%%%%%%%%%%%%%%%%%%%%%%%%%%%%%%%%%%%%%%
\section{Conclusions}
\label{sec-conclude}
%%%%%%%%%%%%%%%%%%%%%%%%%%%%%%%%%%%%%%%%%%%%%

Collective modes in ultrarelativistic ${\cal N} =1$ SUSY QED plasma are essentially the 
same as in the electromagnetic plasma of electrons, positrons and photons. 
Although there are binary processes in supersymmetric plasma, the cross sections of which are 
independent of the momentum transfer, transport properties of ${\cal N} =1$ SUSY QED
plasma are very similar to those of QED one. 

%-----------------------------------------------------------------------
\section*{Acknowledgments}
%-----------------------------------------------------------------------
This work was partially supported by the ESF Human Capital Operational Program under grant
6/1/8.2.1/POKL/2009 and by the Polish Ministry of Science and Higher  Education under grants
N~N202~204638 and 667/N-CERN/2010/0.

%--------------------------------------------------------------------------------------------------------------------------

\end{document}